\documentclass[letterpaper]{article} 
\usepackage{aaai2026}  
\usepackage{times}  
\usepackage{helvet}  
\usepackage{courier}  
\usepackage[hyphens]{url}  
\usepackage{graphicx} 
\usepackage{comment}
\usepackage{natbib}  
\usepackage{caption} 
\usepackage{algorithm}
\usepackage{algorithmic}

\usepackage{amssymb}
\usepackage{amsmath}

\usepackage{xcolor}
\usepackage{tcolorbox}

\usepackage[dvipsnames]{xcolor} 

\usepackage{comment}

\usepackage{newfloat}
\usepackage{listings}
\DeclareCaptionStyle{ruled}{labelfont=normalfont,labelsep=colon,strut=off} 
\floatstyle{ruled}
\newfloat{listing}{tb}{lst}{}
\floatname{listing}{Listing}

\usepackage{booktabs}
\usepackage{multirow}
\title{The Defense Rests: A Courtroom-Inspired
Multi-Agent LLM Framework for Automated Safety
Case Generation}
\author{
    Gerhard Yu, Amanuel Aknaw,  Amir Ali Mohaghegh, Yong Song (Alex) Cao, Alvine B. Belle}
\affiliations{
    Lassonde School of Engineering at York University\\

    4700 Keele St, North York, ON M3J 1P3\\
    Toronto, Canada\\
   gerhardy@yorku.ca, amanuel9@my.yorku.ca, amir1026@my.yorku.ca, alexcao@my.yorku.ca, alvine.belle@lassonde.yorku.ca
}

\usepackage{bibentry}
\usepackage{booktabs}

\begin{document}

\maketitle

\begin{abstract}
Safety cases are widely used to demonstrate that mission-critical systems, such as autonomous vehicles and unmanned aerial systems, can operate safely after deployment. However, manually constructing safety cases is complex, time-consuming, and labor-intensive, which calls for the proposal of automated approaches to support their creation. Inspired by the legal nature of safety case construction, we formalize the safety case creation problem through a courtroom analogy, where stakeholders collaboratively propose, examine, criticize, review, improve and validate safety claims and their supporting evidence. In this analogy, the safety analyst acts as a prosecutor by identifying hazards and safety concerns, while the assurance case developer serves as a defense attorney by constructing and substantiating safety arguments. An orchestrator acts as a judge by coordinating stakeholder interactions, and a review board serves as a jury by assessing argument completeness and persuasiveness. Based on this formalization, we propose \textit{eTrustWare}, a Large Language Model (LLM)-based multi-agent framework for automated safety case generation and review. Our experiments on several safety cases of ML-enabled autonomous driving systems demonstrate the feasibility of coordinated LLM-based agents for generating safety cases.
\end{abstract}

\section{Introduction}
Mission-critical systems are systems whose failure can have significant consequences for human safety, financial loss, or the environment \citep{viger2026evaluating, yu2026llms, sivakumar2024prompting}. Examples include autonomous vehicles, aircraft, medical devices, and critical infrastructure. Assuring their non-functional requirements, such as safety, security, and reliability, is therefore crucial, as these systems' failures can result in severe consequences \citep{ballingall2025safety, finnegan2014security}.

Assurance cases are structured, evidence-based arguments used to demonstrate that a system satisfies a specific property within a defined environment \citep{kelly2007reviewing, maksimov2019survey, meng2025rite}. They consist of claims supported by reasoning, evidence, and explicit system boundaries \citep{annable2025principled, koopman2023ul}. Assurance cases help organizations identify and address risks, demonstrate compliance with regulatory requirements, and communicate confidence in system dependability \citep{koopman2023ul, meng2025rite}.

Existing research has investigated the creation, generation, review, evaluation, formalization, and criticism of assurance cases using AI and large language models (LLMs) \citep{fujiwara2025llm, odu2025automatic, ahlbrecht2025system, viger2024ai, yu2026llms, viger2026evaluating}. However, to the best of our knowledge, only \citep{ikhwantri2026agenticragevaluationframework} has explored Agentic AI for automated assurance-case generation. This limited adoption of Agentic AI constrains the development of efficient, scalable, and intelligent system assurance practices.

Safety cases are a specific form of assurance case focused on system safety \citep{sivakumar2024design}. Inspired by the courtroom analogy \citep{weinstock2009towards, widyasari2026let}, we propose \textit{eTrustWare}, a multi-agent LLM-based framework for automated safety-case generation and review. The framework assigns specialized agents roles analogous to courtroom stakeholders: safety analysts identify hazards, assurance-case developers construct arguments and evidence, an orchestrator plays the role of a judge by coordinating interactions, and a review board evaluates the resulting arguments. Preliminary experiments indicate that \textit{eTrustWare} can generate good safety case drafts that can then be manually refined by experts.

The novelty of our approach lies in combining semi-automated GSN-compliant safety case generation, revision, and scoring, while incorporating human experts' inputs (e.g., safety case patterns) and supporting alignment with industry standards and regulations to better foster the safety of the ML-enabled safety-critical systems under analysis.

Our contributions are four-fold:
\begin{itemize}
    \item We introduce a courtroom-inspired agentic AI approach for the semi-automatic generation of safety cases.
    \item We rely on a dialectic process to iteratively strengthen the generated arguments by supporting their constructive criticism and logical dispute, and by refining them accordingly.
    \item We implement our approach as \textit{eTrustWare}, a tool that supports the generation, revision, scoring, and conversion of generated safety cases from structured prose to GSN.
    \item We evaluate the approach using six safety cases, leveraging various prompting strategies, LLMs, and safety case patterns.
\end{itemize}


\section{Background and Related Work}
\label{sec:backgroung-related-work}

\subsection{Assurance Cases,  Patterns, and Their Notations}

Assurance cases are a well-established structured technique used to document a reasoned, auditable argument supporting that a system meets specific non-functional requirements, such as safety, security and reliability \citep{denney2026advocate, carlan2016integrated}. Assurance cases are structured as a hierarchy of claims, with lower-level claims drawing on concrete evidence and serving as evidence to justify claims higher in the hierarchy \citep{carlan2020enhancing, kelly2007reviewing}. Assurance cases are becoming very popular and are usually used to argue the safety of systems in domains such as automotive, healthcare and aerospace \citep{buysse2025safe, carlan2016integrated, sivakumar2024design, maksimov2019survey}. 

Different assurance cases address different non-functional requirements \citep{delavara2023assurance, finnegan2014security}. This work focuses on safety cases, which demonstrate that systems are acceptably safe for intended operations and are particularly important for mission-critical systems where failures may cause significant harm \citep{borg2023ergo, gauerhof2020assuring, buysse2025safe}.

Several notations allow representing assurance cases. This work focuses on the very popular GSN (Goal Structuring Notation) and GSN-compliant structured prose, which clearly represent assurance relationships for both humans and machines \citep{maksimov2019survey, odu2025smartgsn}.

Assurance case patterns provide reusable structures for constructing arguments by capturing recurring relationships among claims, assumptions, contexts, and evidence \citep{carlan2020enhancing, finnegan2014security}. When used to create assurance cases, they improve their quality. 

\subsection{LLMs (Large Language Models) and Agentic AI}

As \citet{hou2024large} explain, LLMs are AI models trained on massive textual datasets and able to understand and generate human language, learning linguistic patterns, semantics, and context. They support reasoning, decision-making, and automation across applications such as generation, retrieval, coding, testing, and documentation. Their reliance of various prompting strategies, including zero-shot, few-shot, and chain-of-thought usually improve the quality of their outputs \citep{mohajer2024effectiveness, odu2025automatic, yu2026llms}.

LLM-based agents are (semi-)autonomous systems that use LLMs as their central reasoning component to perceive their environment, make decisions, and execute actions \cite{he2025llm}. LLM-based agents yield several benefits when it comes to the automation of software engineering tasks: they can autonomously plan and execute multi-step workflows, interact with development tools, generate and modify code, detect and fix defects, and assist with testing and documentation \citep{he2025llm, zeng2026experience}. This agentic capability enables LLMs to move beyond isolated task completion toward more autonomous and context-aware software engineering processes \citep{zeng2026experience}.

\subsection{Related Work on LLM-based System Assurance}

There is a rich body of work on LLM-based system assurance (e.g., \citep{diemert2025balancing, herd2026subjective, ghahremani2024avoiding, muram2023attest, fujiwara2025llm, odu2025automatic, mohamad2023cascade, buysse2025safe, viger2024ai, gohar2024codefeater, yu2026llms, sivakumar2024prompting, di2024mmint, odu2025smartgsn, viger2026evaluating,  viger2023supporting, murugesan2024automating, shahandashti2024using}). \citet{belle2025preliminary} surveyed some of them. LLM-based system assurance approaches focus on various activities including assurance case creation (e.g., \citep{odu2025automatic, sivakumar2024design}), defeater identification (e.g., \citep{shahandashti2024using, viger2024ai, gohar2024codefeater, viger2026evaluating}), assurance case review (e.g., \cite{yu2026llms}), verification, formalization, semantic analysis, and assessment. However, most approaches focus on assurance case creation and defeater identification. Our focus is on assurance case creation, and more specifically on the semi-automatic safety case creation.

Existing LLM-based assurance case approaches usually rely on models from the GPT series to generate assurance cases complying with notations such as GSN, EA (Eliminative Argumentation), CAE (Claim-Argument-Evidence), or logical programming. For instance, \citet{sivakumar2024prompting} used GPT-4 to generate safety cases, while \citet{odu2025automatic} used GPT-4 Turbo and GPT-4o to generate assurance cases from predicate-based assurance case patterns. This approach was implemented as a tool called \textit{SmartGSN} \citep{odu2025smartgsn} and applied to the automotive sector \citep{odu2025llm}. Likewise, \citet{fujiwara2025llm} proposed a GPT-based approach for generating assurance cases and mitigations for AI-system threats. These approaches often employ prompting strategies such as zero-shot, one-shot, role-based prompting, and Chain-of-Thought when generating assurance cases. However, hallucinations, non-deterministic outputs, limited explainability, and insufficient real-world validation often constrain the reliability and practical applicability of these approaches.

To the best of our knowledge, only one approach combines RAG (Retrieval-Augmented Generation) and agentic AI for assurance case generation. More specifically,  \citet{ikhwantri2026agenticragevaluationframework} proposed an Agentic RAG  framework that represents assurance cases using CAE and grounds arguments in organisation-specific technical documentation retrieved from controlled local repositories. In our work, we also use Agentic RAG, but differently from Ikhwantri et al., we introduce a multi-agent, courtroom-inspired framework in which prosecutor, defense attorney, judge, and jury agents collaboratively construct, challenge, validate, and review safety claims and evidence. This enables reliance on a dialectic process \cite{gsnstandard} that promotes evidence-grounded generation alongside independent critical review within a unified framework.

\section{Proposal of  \textit{eTrustware}: A Courtroom-Inspired Multi-Agent LLM Framework
for Automated Safety Case Generation}
\label{sec:proposal}

\subsection{Proposed Approach: Mapping the Courtroom Analogy to the Generation/Review of Safety Cases}

To address the complexities of automated reasoning and the limitations of single-agent systems, \citet{widyasari2026let} recently proposed an approach that adapts a multi-agent, courtroom-inspired framework to perform automated source-code vulnerability detection, where role-specific agents debated potential security risks with each other before a judge agent delivers a final classification and explanation. Their results show that their courtroom-inspired multi-agent approach outperforms both the
single-agent and traditional multi-agent baselines.

Conventional automated systems rely on one LLM, limiting perspective diversity and increasing hallucinations, overconfidence, and errors, which can compromise safety arguments in mission-critical systems. These limitations become particularly significant when reasoning about mission-critical systems, where an incorrect assumption or unsupported claim can undermine the validity of the resulting safety argument. 

The courtroom analogy allows us to address these challenges by distributing the safety case creation and review tasks across multiple distinct role-specific agents. By mimicking a legal trial, we define a new framework called \textit{eTrustWare} that enforces a structured, adversarial discussion. The four distinct courtroom-inspired roles it supports are mainly: The \textit{Prosecutor}, the \textit{Defense Attorney}, the \textit{Judge}, and the \textit{Jury}. The prosecutor is responsible for investigating the evidence, establishing the facts, and building a case against the system. The defense attorney is tasked with advocating for integrity in the system, typically by countering the prosecutor's claims. There are some cases where a certain safety case flaw is valid; in such cases, the defense attorney has to acknowledge the issue and propose mitigation strategies. The judge is responsible for upholding fairness in the courtroom. This agent must filter out bias and act in an objective manner. The jury is responsible for weighing all evidence and helps with the delivery of the final verdict. This structured process could help reduce wrongful conclusions and force a deeper and more iterative analysis in complex and/or debatable argumentation structures.

In the safety case domain, this analogy is highly applicable, because safety cases are fundamentally structured arguments in which claims must be justified through supporting evidence and reasoning. Similar to a legal case, the strengths of a safety cases depends not only on the evidence presented, but also if that evidence adequately supports the claims and whether the reasoning connecting them is sound and complete. Thus, the courtroom creates a natural mechanism for examining the validity of these arguments from multiple perspectives. By having one role construct and defend the claim, and another challenge its claims, unsupported evidence, assumptions, and potential weaknesses, the approach can help expose deficiencies that may be overlooked by a single agent.

\begin{figure}[htbp]
    \centering 
    \includegraphics[width=0.5\textwidth]{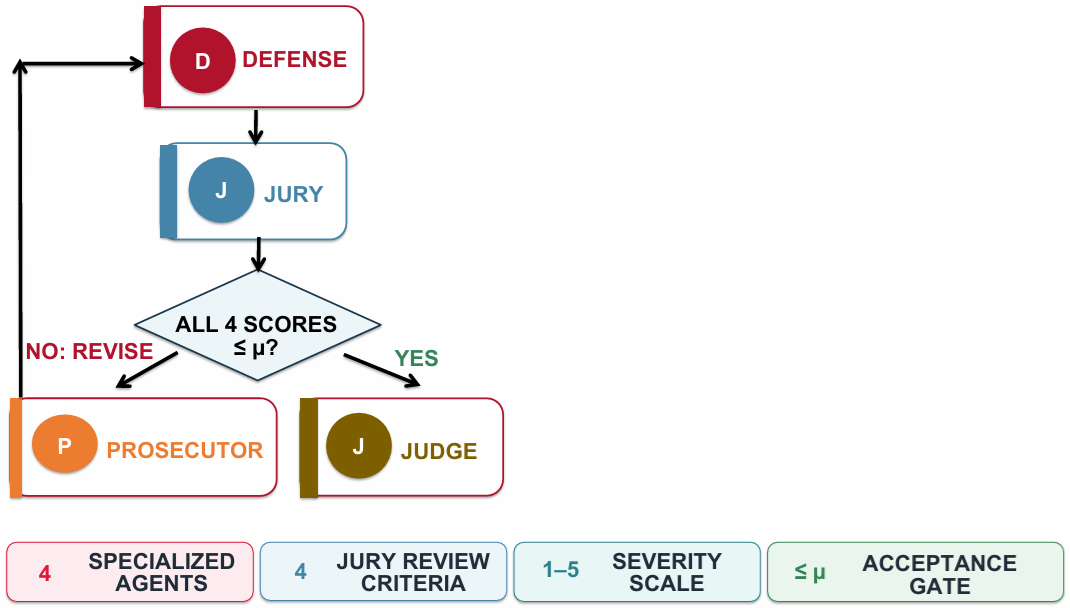} 
    \caption{\textit{eTrustWare} agentic workflow}
    \label{fig:agentic_workflow} 
\end{figure}

We have adapted the mock-court approach utilizing the philosophy it conveys. More specifically, our Defense agent is responsible for generating the initial safety case utilizing safety case patterns. The Prosecutor agent critically examines the generated cases by identifying unsupported claims, insufficient evidence, inconsistencies, and other potential weaknesses. The Defense agent is then responsible for responding to this feedback, whether it is to justify, address the limitations, or refine the argument. The Jury agent evaluates the argument on four different review criteria that are recommended by the GSN standard \cite{gsnstandard}: argument comprehension, expressive sufficiency, well-formedness, and resistance to argument criticism and defeat. These four criteria are scored on a scale of 1-5, and the total score alongside the average is weighted to decide if the case goes back to the Defense agent or the jury. If the stores have met the standard, the Judge agent is then responsible for doing a final evaluation of the generated safety case. Through this adaptation, the courtroom analogy provides a structured workflow where safety cases are generated, revised, and evaluated across multiple perspectives, addressing the previous limitations of a single-agent system. Figure~\ref{fig:agentic_workflow} illustrates this workflow.

\subsection{Description of the Features of \textit{eTrustWare}}
\textit{eTrustWare} is meant to support the semi-automatic generation of safety cases. So, it aims at automatically generating safety cases that can then be manually refined by human experts (e.g., safety case developers).

\textit{eTrustWare} is also designed to support various non-functional requirements. More specifically, it aims to be secure, scalable, web-accessible, and maintainable. To support these requirements, \textit{eTrustWare} adopts a modular, tiered architecture that separates its React-based user interface, FastAPI backend, LLM-based agents, knowledge-retrieval capabilities through RAG, and a Google Firebase persistence tier. Authentication, authorization, and database access rules protect system and user data, while cloud-based storage and retrieval services help with scalability. The web-based interface allows for remote access, and the microservice architecture aids in maintenance, testing, debugging, and future additions.

\subsection{Description of the Architecture of \textit{eTrustWare}}

The architecture of \textit{eTrustWare} consists of four tiers: 

\textbf{The Presentation Tier:} It is the tier that is responsible for displaying the data received from the other tiers and allowing users to interact with\textit{eTrustWare} features, such as authentication, creating and editing new assurance cases using GSN diagrams and structured prose, and exporting assurance cases. Furthermore, this tier sends requests to the other tiers to generate assurance cases and persist user data.

\textbf{The Logic (agentic) Tier:} This is the tier where the various agents collaborate to generate safety cases in structured prose based on the provided prompts.

\textbf{Knowledge Tier}: It retrieves relevant safety standards and regulations for evidence-based reviews.  

\textbf{Persistence Tier}: This tier stores safety case versions, agent interactions, scores, and audit trails.

\textit{eTrustWare} is built atop \textit{SmartGSN} \cite{odu2025automatic} and adapts some of its publicly available prompts to generate safety cases. \textit{SmartGSN} is an LLM-based tool allowing to semi-automatically create assurance cases from assurance case patterns supplied in the prompts of its LLMs. Unlike \textit{SmartGSN}, \textit{eTrustWare} uses an agentic RAG approach based on the courtroom analogy. Thanks to that analogy and unlike existing work (e.g., \cite{ikhwantri2026agenticragevaluationframework}), \textit{eTrustWare} is able to use a dialectic approach \cite{gsnstandard} to review and criticize the safety case it generates, and to iteratively propose improvements to refine it.

\subsection{Description of the RAG Pipeline}


To implement our RAG pipeline, we used the best practices recommended in the literature (e.g., \citep{lewis2020retrieval, brown2025systematic}). More specifically, we implemented a conventional (naïve) RAG approach based on dense vector similarity retrieval. Our RAG corpus consists of safety and assurance standards, more specifically AMLASv1.1 \cite{hawkins2021guidance} and UL4600 \cite{koopman2023ul}. We selected these documents because they provide domain-specific guidance relevant to the assurance of autonomous driving systems. The documents are processed by extracting their text and dividing it into chunks of 800 characters with a 100-character overlap. Each chunk is then converted into vectors using OpenAI's text-embedding-3-small model and stored in a Pinecone vector database using cosine similarity.

The RAG pipeline is integrated into the Prosecutor and Judge agents, which main purpose includes reviewing the generated assurance case. When these agents require additional safety-assurance knowledge, their queries are used to retrieve the two most semantically relevant chunks from the standards corpus. By grounding the critical agents in relevant standards-based information, the RAG pipeline provides domain-specific knowledge that strengthens the review process and supports the generation of well-founded safety case arguments.

\section{Implementation of \textit{eTrustWare}}
\label{sec:implementation}


\subsection{Presentation Tier Implementation} 
This tier consists of a single webpage with multiple tabs to allow for coherent navigation. We implemented this tier with React, TypeScript, Tailwind CSS, ReactFlow, and Shadcn as a component library to provide users with an interactive interface to log in and sign up, along with accessing the core features. Users can automatically generate a new safety case, edit it, and save it. Additionally, this tier allows users to create a manual GSN diagram from scratch or generate one from the one automatically created.

\subsection{FastAPI Gateway Implementation}

The gateway provides a communication interface between the presentation tier and the underlying \textit{eTrustware} components. We implemented this tier using FastAPI, with Uvicorn serving as the application server to handle incoming HTTP requests. The gateway contains endpoints where users could send and receive requests allowing them to generate assurance cases and interact with the system, forwarding requests to the logic tier and return results back to the presentation tier. This tier provides a consistence interface between the presentation tier and the logic, knowledge, and data tier while keeping their internal implementations of these components separated from the client.

\subsection{Logic Tier (agentic tier) Implementation}

The logic tier implements \textit{eTrustware}'s courtroom inspired multi agent architecture using a collection of specialized agents. The tier consists of the Defense agent, Prosecutor agent, Jury agent, and Judge agent, with each agent performing a distinct role in the safety case generation and evaluation process. The Prosecutor and Judge agents are additionally connected to the knowledge tier to retrieve relevant information from safety-assurance standards during critical evaluation, while the resulting agent outputs and execution information are passed through the persistence mechanisms as required.

\subsection{The Knowledge Tier Implementation}

The knowledge tier provides the agents with access to domain-specific safety assurance knowledge through the integrated RAG infrastructure described earlier. The Prosecutor and Judge agents interact with this tier when standards-based information is required during their evaluation process, allowing relevant knowledge from the corpus to be incorporated and to help justify their reasoning.

\subsection{ Data Tier (persistence tier) Implementation}

The data tier is implemented using Firebase Firestore to provide persistent storage for \textit{eTrustware}'s data. Firestore's document based structure is used to organize and persist information associated with users, safety cases, generated content, graph nodes, and other data required throughout the system's workflow. The logic tier interacts with Firestore to store and retrieve this information as required, allowing generated safety cases and related data to remain available across sessions and keeping data management separated from the rest of the system.

\section{Setup of the Preliminary Assessment}
\label{sec:experimental-setup}
\subsection{Research Questions}
Our experiments investigated three research questions (RQs):

\textbf{RQ1}: Which LLM is the best suited to serve as the core reasoning engine for the \textit{eTrustWare} multi-agent framework?

\textbf{RQ2}: What is the impact of supplying patterns in the prompts of \textit{eTrustWare} when generating safety cases?

\textbf{RQ3}: What is the impact of prompting strategies on the quality of the safety cases generated by the \textit{eTrustWare}?


\subsection{Description of the Preliminary Experiments}
\label{subsec:experiments_description}

To answer our research questions, we performed the following six experiments. 1) The \textbf{Zero-shot (ZS-Exp) experiment} provides only a brief prompt describing the topic and the relevant tool or technology, without examples, patterns, or additional guidance; 2) the \textbf{Zero-shot + CoT (ZS-CoT-Exp)} adds step-by-step instructions that specify goal order, claim structure, conditions, connections, strategies, domains, and solutions for each subgoal; 3) the \textbf{Zero-shot + CoT + Patterns (ZS-CoT-Pat-Exp)} further adds case-specific design patterns with stricter guidance on the number of goals and solutions, decorators, and system context; 4) The \textbf{One-shot (OS-Exp)} experiment adds a high-quality input-output example involving similar technology or the same project; 5) the \textbf{One-shot + CoT (OS-CoT-Exp)} combines this example with the CoT instructions and structural guidance; and 6) the \textbf{One-shot + CoT + Patterns (OS-CoT-Pat-Exp)} combines all previous elements with unique, stricter design patterns that guide goals, solutions, decorators, and system-specific context.

Note that when running a given experiment using a given LLM as the inference mechanism of \textit{eTrustWare}, the latter generates a safety case in structured prose. When needed, we then manually clean this safety case to remove hallucinated symbols. We then use \textit{eTrustWare} to automatically convert the resulting safety case into a graphical format, i.e., into a GSN diagram.

\subsection{Dataset Description and Representation in GSN and Structured Prose}

In our experiments, we use a dataset consisting of six GSN-compliant safety cases and the patterns used to manually create them. Table \ref{tab:dataset}
 describes the characteristics of these safety cases and the associated patterns. Although all six safety cases focus on ML-enabled autonomous driving systems, they cover many different safety-related scenarios, including trajectory prediction, collision avoidance, natural-language instruction controls, pedestrian automatic braking, automated valet parking, and self-adapting abilities during runtime.

\begin{table*}[htbp]
    \centering
    \caption{Characteristics of safety cases and patterns in dataset}
    \begin{tabular}{@{}llllll@{}}
    \toprule
        \textbf{Safety case/system} & \textbf{\#Pat. decorators} & \textbf{\#Pat. placeholders} & \textbf{\#Pat. elements} & \textbf{\#GT elements} & \textbf{\#GT relationships} \\
    \midrule
        Baidu Apollo & 37 & 22 & 49 & 38 & 59\\
        Quanser QCar & 37 & 22 & 49 & 24 & 23\\
        SimLingo & 23 & 19 & 25 & 37 & 36\\
        SMIRK & 45 & 54 & 74 & 80 & 103\\
        Automated Valet Parking & 0 & 0 & 22 & 22 & 22\\
        F1TENTH & 0 & 14 & 11 & 32 & 31\\
    \bottomrule
    \end{tabular}
    \label{tab:dataset}
\end{table*}

As ground truths, we used these six manually created safety cases represented as textual prose. These safety cases are publicly available in the following references \cite{odu2025llm, sivakumar2024design,  yu2026safety, borg2023ergo, sorokin2024towards, nearchou2023assurance}. 

For our one-shot experiments, we used the manually authored safety case for the continuous validation of a Level 4 automated shuttle bus under ISO guidelines \cite{kodama2024} as the one-shot example. 
In the \textbf{One-shot + CoT + patterns} experiment, the formalized pattern used to manually create this safety case was also supplied as part of the one-shot example.

\subsection{Description of the LLMs and their Settings}

\textit{eTrustWare} uses an LLM to support the inference mechanism of its various agents. This LLM allows these agents to make automated decisions. In our experiments, we tried different LLMs on \textit{eTrustWare} to decide which of these LLMs is the best to support decision-making in this agentic system. The LLMs we used in our experiments are the following: GPT-5.5, Claude Fable-5, and DeepSeek-v4-Pro. The temperature of the LLMs was set at 0.2. This relatively low temperature allowed for reduced variability in the generated responses and promoted more consistent and reproducible decision-making across experimental runs. Table \ref{tab:llm-settings} describes the characteristics of these LLMs.

\begin{table}[htbp]
  \centering
  \caption{Settings and characteristics of the evaluated LLMs}
  \label{tab:llm-settings}
  \footnotesize
  \begin{tabular}{@{}lllll@{}}
    \toprule
    \textbf{Model} & \textbf{Org.} & \textbf{Context} & \textbf{Max Out}  \\
    \midrule
    GPT-5.5         & OpenAI    & $\sim$1.05M & 128K        \\
    Claude Fable-5  & Anthropic & 1M          & 128K       \\
    DeepSeek-V4-Pro & DeepSeek  & 1M          & up to 384K \\
    \bottomrule
  \end{tabular}
  \vspace{-0.5em}
\end{table}


\subsection{Description of \textit{eTrustWare} Settings}

As shown in figure~\ref{fig:agentic_workflow}, \textit {eTrustWare} evaluates the safety cases it generates by relying on four review criteria: expressive sufficiency, well-formedness, argument comprehension, and resistance to criticism and defeat. The Jury agent scores each criterion on a discrete 1--5 scale, where 1 represents the best and 5 the worst outcome. Decimal scores are avoided because finer-grained scoring provides limited benefit while potentially increasing execution and iteration times. The four scores are averaged and considered alongside the individual criterion scores when applying the acceptance gate. The gate requires every criterion to achieve a score of at most $\mu$, currently set to 2. Safety cases meeting this threshold proceed to the Judge agent, which delivers the result to the user. Cases failing any criterion are returned to the Prosecutor agent for criticism and subsequently to the Defense agent for revision. The process currently permits three iterations before force-quitting. Manual evaluation against ground truth demonstrated that $\mu=2$ produced the best results. 

\subsection{Description of the Assessment Metrics}
In accordance with the literature \cite{odu2025automatic, sivakumar2024prompting}, to assess our experimental results, we relied on the following metrics that allow us to evaluate the structural and semantic correctness of the safety  cases \textit{eTrustWare} generated:

\begin{itemize}
\item \textbf{Tree Edit Distance (TED)}: TED measures structural differences between generated and ground-truth safety cases. To compute this metric, we represent each safety case as a rooted, ordered tree, with GSN elements as nodes and support relationships as edges. TED counts the minimum node insertions, deletions, and relabellings required to transform the generated tree into the ground truth. We compute TED using the Zhang--Shasha algorithm through Python's \textit{zss} library. A value of zero indicates identical structures, while larger values indicate greater structural differences.

\item \textbf{Cosine Similarity}: Cosine similarity measures semantic similarity between the generated safety case and its ground-truth \cite{odu2025automatic, sivakumar2024design}.Although cosine similarity generally
ranges from -1 to 1, the use of non-negative word-frequency vectors
restricts the range in this study from 0 to 1. A score of zero indicates
that the two safety cases share no common terms. A score of one indicates that their word-frequency vectors have the same direction and
therefore display maximum similarity. Values closer to one indicate
greater semantic similarity, whereas values closer to zero indicate less
similarity. To compute this metric, we relied on the \textit{CountVectorizer} Python class from
the \textit{sklearn} library.
\end{itemize}


\section{Preliminary Results} 
\label{sec:prelim-results}

\subsection{(RQ1): On the Selection of the Right LLM for the \textit{eTrustWare} Multi-Agent Framework}

Tables  \ref{tab:cosine_similarity} and \ref{tab:tree_edit_distance}  report the results of our experiments. Both tables indicate that Fable-5 is the most suitable LLM to serve as the core reasoning engine for the \textit{eTrustWare} multi-agent framework. Across the six safety cases and prompting configurations, Fable-5 generally achieves lower Tree Editing Distance (TED), indicating greater structural similarity to the reference safety cases, while simultaneously maintaining comparatively high cosine similarity. Averaged across all experiments, Fable-5 achieves an approximate TED of 68, compared with 75 for DeepSeek-v4-pro and 128 for GPT-5.5. Its average cosine similarity is also the highest, at approximately 0.57, compared with 0.49 for DeepSeek-v4-pro and 0.47 for GPT-5.5. Although DeepSeek-v4-pro occasionally achieves the highest individual scores, Fable-5 provides the most consistent overall compromise between structural and semantic quality across safety cases.

\begin{table*}[h!]
\centering
\caption{Cosine Similarity Results}
\label{tab:cosine_similarity}
\small
\renewcommand{\arraystretch}{1.08}
\setlength{\tabcolsep}{6pt}
\begin{tabular}{llccc}
\toprule
\textbf{Safety Case} & \textbf{Prompting technique/Experiments} & \textbf{GPT-5.5} & \textbf{Fable-5} & \textbf{DeepSeek-v4-pro} \\
\midrule
\multirow{6}{*}{Baidu Apollo}
 & Zero-shot                       & 0.42 & 0.48 & 0.37 \\
 & Zero-shot + CoT                 & 0.40 & 0.37 & 0.49 \\
 & Zero-shot + CoT + patterns      & 0.64 & \textbf{0.76} & 0.77 \\
 & One-shot                        & 0.42 & 0.43 & 0.54 \\
 & One-shot + CoT                  & 0.45 & 0.46 & 0.44 \\
 & One-shot + CoT + patterns       & \textbf{0.72} & 0.55 & \textbf{0.82} \\
\midrule
\multirow{6}{*}{Quanser QCar}
 & Zero-shot                       & 0.49 & 0.55 & 0.39 \\
 & Zero-shot + CoT                 & 0.45 & 0.66 & 0.55 \\
 & Zero-shot + CoT + patterns      & 0.73 & 0.73 & 0.74 \\
 & One-shot                        & 0.50 & 0.57 & 0.49 \\
 & One-shot + CoT                  & 0.53 & 0.55 & 0.59 \\
 & One-shot + CoT + patterns       & \textbf{0.73} & \textbf{0.79} & \textbf{0.80} \\
\midrule
\multirow{6}{*}{SimLingo}
 & Zero-shot                       & 0.49 & 0.61 & 0.52 \\
 & Zero-shot + CoT                 & 0.44 & 0.58 & 0.32 \\
 & Zero-shot + CoT + patterns      & 0.61 & \textbf{0.75} & \textbf{0.69} \\
 & One-shot                        & 0.44 & 0.43 & 0.38 \\
 & One-shot + CoT                  & 0.48 & 0.40 & 0.41 \\
 & One-shot + CoT + patterns       & \textbf{0.63} & 0.62 & 0.68 \\
\midrule
\multirow{6}{*}{SMIRK}
 & Zero-shot                       & 0.44 & 0.60 & 0.36 \\
 & Zero-shot + CoT                 & 0.43 & 0.64 & 0.38 \\
 & Zero-shot + CoT + patterns      & 0.72 & 0.75 & 0.74 \\
 & One-shot                        & 0.44 & 0.46 & 0.50 \\
 & One-shot + CoT                  & 0.45 & 0.63 & 0.54 \\
 & One-shot + CoT + patterns       & \textbf{0.77} & \textbf{0.78} & \textbf{0.84} \\
\midrule
\multirow{6}{*}{\shortstack[l]{Automated Valet\\Parking}}
 & Zero-shot                       & 0.42 & 0.41 & 0.33 \\
 & Zero-shot + CoT                 & 0.38 & 0.60 & 0.62 \\
 & Zero-shot + CoT + patterns      & \textbf{0.53} & 0.70 & 0.74 \\
 & One-shot                        & 0.43 & 0.68 & 0.57 \\
 & One-shot + CoT                  & 0.38 & 0.69 & 0.68 \\
 & One-shot + CoT + patterns       & 0.45 & \textbf{0.79} & \textbf{0.84} \\
\midrule
\multirow{6}{*}{\shortstack[l]{F1TENTH\\Racing Car}}
 & Zero-shot                       & 0.30 & \textbf{0.65} & 0.33 \\
 & Zero-shot + CoT                 & 0.35 & 0.46 & \textbf{0.51} \\
 & Zero-shot + CoT + patterns      & 0.68 & 0.56 & 0.09 \\
 & One-shot                        & 0.50 & 0.21 & 0.14 \\
 & One-shot + CoT                  & 0.37 & 0.20 & 0.16 \\
 & One-shot + CoT + patterns       & \textbf{0.75} & 0.21 & 0.19 \\
\bottomrule
\end{tabular}
\end{table*}

\begin{table*}[h!]
\centering
\caption{Tree Editing Distance (TED) Results}
\label{tab:tree_edit_distance}
\small
\renewcommand{\arraystretch}{1.08}
\setlength{\tabcolsep}{6pt}
\begin{tabular}{llccc}
\toprule
\textbf{Safety Case} & \textbf{Prompting technique/Experiment} & \textbf{GPT-5.5} & \textbf{Fable-5} & \textbf{DeepSeek-v4-pro} \\
\midrule
\multirow{6}{*}{Baidu Apollo}
 & Zero-shot                       & 215 & 89 & 86 \\
 & Zero-shot + CoT                 & 115 & 63 & 74 \\
 & Zero-shot + CoT + patterns      & 122 & 72 & 83 \\
 & One-shot                        & 173 & 63 & 77 \\
 & One-shot + CoT                  & 170 & 63 & 89 \\
 & One-shot + CoT + patterns       & \textbf{74} & \textbf{60} & \textbf{72} \\
\midrule
\multirow{6}{*}{Quanser QCar}
 & Zero-shot                       & 153 & 88 & 46 \\
 & Zero-shot + CoT                 & 116 & 27 & 39 \\
 & Zero-shot + CoT + patterns      & 66 & 98 & \textbf{27} \\
 & One-shot                        & 117 & 27 & 82 \\
 & One-shot + CoT                  & 82 & 29 & 41 \\
 & One-shot + CoT + patterns       & \textbf{62} & \textbf{24} & 49 \\
\midrule
\multirow{6}{*}{SimLingo}
 & Zero-shot                       & 132 & 87 & 41 \\
 & Zero-shot + CoT                 & 108 & 39 & 51 \\
 & Zero-shot + CoT + patterns      & 121 & 50 & 42 \\
 & One-shot                        & 121 & 39 & 87 \\
 & One-shot + CoT                  & 128 & 38 & 44 \\
 & One-shot + CoT + patterns       & \textbf{72} & \textbf{37} & \textbf{37} \\
\midrule
\multirow{6}{*}{SMIRK}
 & Zero-shot                       & \textbf{118} & 130 & \textbf{107} \\
 & Zero-shot + CoT                 & 125 & \textbf{104} & \textbf{107} \\
 & Zero-shot + CoT + patterns      & \textbf{118} & 175 & 120 \\
 & One-shot                        & 230 & 105 & 141 \\
 & One-shot + CoT                  & 187 & 106 & 108 \\
 & One-shot + CoT + patterns       & \textbf{118} & 105 & 118 \\
\midrule
\multirow{6}{*}{\shortstack[l]{Automated Valet\\Parking}}
 & Zero-shot                       & 89 & 111 & 152 \\
 & Zero-shot + CoT                 & 86 & 29 & 59 \\
 & Zero-shot + CoT + patterns      & 129 & 43 & 57 \\
 & One-shot                        & 138 & 30 & 68 \\
 & One-shot + CoT                  & 182 & \textbf{28} & 78 \\
 & One-shot + CoT + patterns       & \textbf{72} & 37 & \textbf{38} \\
\midrule
\multirow{6}{*}{\shortstack[l]{F1TENTH\\Racing Car}}
 & Zero-shot                       & 118 & 124 & 48 \\
 & Zero-shot + CoT                 & 110 & 35 & 89 \\
 & Zero-shot + CoT + patterns      & 125 & 54 & 53 \\
 & One-shot                        & 117 & 34 & 136 \\
 & One-shot + CoT                  & 98 & 37 & \textbf{44} \\
 & One-shot + CoT + patterns       & \textbf{94} & \textbf{33} & 54 \\
\bottomrule
\end{tabular}
\end{table*}

\subsection{(RQ2): Impact of the use of patterns on \textit{eTrustWare} Performance}

The results in Tables \ref{tab:cosine_similarity} and \ref{tab:tree_edit_distance} show that including patterns as LLM inputs (i.e. prompting an LLM to use a pattern to generate a safety case) usually improves the quality of the safety cases \textit{eTrustWare} generates. Compared with corresponding configurations without patterns, both TED and cosine similarity frequently improve when patterns are supplied in LLM prompts. For instance, in the \textbf{One-shot + CoT} experiment, adding patterns reduces TED substantially for GPT-5.5 in Baidu Apollo (170 to 74), Quanser QCar (82 to 62), and SimLingo (128 to 72), while cosine similarity increases from 0.45 to 0.72, 0.53 to 0.73, and 0.48 to 0.63, respectively. Similar improvements are observed for Fable-5 and DeepSeek-v4-pro across several safety cases. 

\subsection{(RQ3): Impact of the use of various prompting strategies on \textit{eTrustWare} Performance}

The results in Tables \ref{tab:cosine_similarity} and \ref{tab:tree_edit_distance} also show that prompting strategy has a substantial impact on the generated safety case quality, with the \textbf{One-shot + CoT + patterns} experiment generally producing the best results. This experiment usually yields the lowest TED and highest cosine similarity, particularly for GPT-5.5 and Fable-5. For instance, GPT-5.5 achieves its lowest TED for Baidu Apollo (74), Quanser QCar (62), SimLingo (72), Automated Valet Parking (72), and F1TENTH Racing Car (94) under this strategy. It also obtains its highest cosine similarity for these safety cases, except for Automated Valet Parking. Likewise, Fable-5  achieves its best or near-best results under pattern-enhanced prompting. However, performance varies across models and safety cases, suggesting that prompting effectiveness is model- and context-dependent. 

\section{Discussion}
\label{sec:discussion}

\textit{eTrustWare} demonstrates strong potential for safety case generation using patterns, structured prompts, chain-of-thought, and one-shot techniques. However, the safety cases it generates do not yet consistently meet industry standards. Its courtroom-based iterative process improves safety cases through defense, jury evaluation, and prosecutor critiques, although safety case revisions may sometimes deviate from pattern structures or introduce additional GSN elements that do not exist in the ground truth.  Manually refining the safety cases that \textit{eTrustWare} yields therefore remains crucial. 

Our experiments focused on commercial LLMs. Still, high-quality operation is also costly, with approximately CAD $0.20–$2 per generation due to LLM token usage. Future work could integrate free or local LLMs to reduce costs and improve privacy. Scalability may improve with more performant LLMs, refined agent prompts, additional agents, user intervention, and real-time collaborative editing for large safety cases.

\section{Threats to Validity}
\label{sec:validity-threats}

A potential threat to external validity is the generalization of our results beyond the analyzed safety cases, safety standards, LLMs, and experimental configurations used in our evaluation. In particular, \textit{eTrustware} was evaluated in the context of autonomous system safety cases because, in the automotive sector, safety is paramount. As a consequence, the observed performance may differ when \textit{eTrustware} is applied to other safety-critical domains, types of assurance cases, or assurance standards. Similarly, results obtained with the selected LLM and prompting strategies may not necessarily generalize to other models or newer versions of LLMs. To mitigate this threat, we evaluated multiple experimental configurations and used established safety assurance standards relevant to autonomous driving systems. Nevertheless, additional experiments using larger and more diverse safety cases from additional domains, standards, LLMs, and prompting strategies are required to assess the generalization of our findings.

\section{Conclusion and Future Work}
\label{sec:conclusion}

In this paper, we introduced \textit{eTrustWare}, a courtroom-inspired multi-agent LLM framework for generating and evaluating autonomous-vehicle safety cases. Experiments on \textit{eTrustWare} showed that Claude Fable-5 performs best, while patterns, chain-of-thought prompting, and one-shot examples improve the generated safety cases structural and semantic correctness. 

Future work will extend \textit{eTrustWare} through real-time human-agent collaboration, will allow deploying it, and will aim at validating it using additional evaluation metrics and a larger dataset covering diverse application domains such as healthcare, nuclear, and aerospace, alongside standards such as ISO 26262. Manual expert assessments will examine potential LLM non-determinism and hallucinations.

\section*{Acknowledgments}

This work was undertaken thanks in part to funding from the Connected Minds program, supported by Canada First Research Excellence Fund, grant \#CFREF-2022-00010. This work was also partially supported by the NSERC (Natural Sciences and Engineering Research Council of Canada) USRA program, by an NSERC Discovery grant, and by the Lassonde School of Engineering (at York University) LURA program. Some features of \textit{eTrustWare} are inspired by \textit{SmartGSN} \cite{odu2025smartgsn}. We therefore want to thank the original developers of \textit{SmartGSN}, namely: Oluwafemi Odu, Daniel Méndez Beltrán, Emiliano Berrones Gutiérrez, Melika Sherafat, Mhd Ryiad Alelaioy, Ursanne Kengne, Alexis Lavigne, and Ethan Ngo.







\bibliography{atracc-2026_etrustware_references}

@article{widyasari2026let,
  title={Let the trial begin: A mock-court approach to vulnerability detection using llm-based agents},
  author={Widyasari, Ratnadira and Weyssow, Martin and Irsan, Ivana Clairine and Ang, Han Wei and Liauw, Frank and Ouh, Eng Lieh and Shar, Lwin Khin and Kang, Hong Jin and Lo, David},
  journal={International Conference in Software engineering},
  year={2026}
}

@inproceedings{mohajer2024effectiveness,
  title={Effectiveness of chatgpt for static analysis: How far are we?},
  author={Mohajer, Mohammad Mahdi and Aleithan, Reem and Harzevili, Nima Shiri and Wei, Moshi and Belle, Alvine Boaye and Pham, Hung Viet and Wang, Song},
  booktitle={Proceedings of the 1st ACM International Conference on AI-Powered Software},
  pages={151--160},
  year={2024}
}

@misc{ikhwantri2026agenticragevaluationframework,
  title={An Agentic RAG and Evaluation Framework for Assurance Case Generation: Industrial Use Case for the EU Cyber Resilience Act Compliance}, 
  author={Fariz Ikhwantri and Iker Lasa Ojanguren and Dusica Marijan and Pavlos Kosmides},
  year={2026},
  month={August},
  eprint={2608.19509},
  archivePrefix={arXiv},
  primaryClass={cs.SE},
  doi={10.48550/arXiv.2608.19509},
  url={https://doi.org/10.48550/arXiv.2608.19509}
}

@inproceedings{meng2025rite,
  title={RITE: An IDE for Assurance and Certification of Software and Systems},
  author={Meng, Baoluo and Paul, Saswata and Varanasi, Sarat Chandra and Alexander, Christopher and Mertens, Eric and Moitra, Abha and Siu, Kit and Durling, Michael},
  booktitle={2025 AIAA DATC/IEEE 44th Digital Avionics Systems Conference (DASC)},
  pages={1--10},
  year={2025},
  organization={IEEE}
}

@article{buysse2025safe,
  title={Safe autonomous systems in a changing world: Operationalising dynamic safety cases},
  author={Buysse, Laure and Habli, Ibrahim and Vanoost, Dries and Pissoort, Davy},
  journal={Safety Science},
  volume={191},
  pages={106965},
  year={2025},
  publisher={Elsevier}
}

@inproceedings{denney2026advocate,
  title={AdvoCATE: The Assurance Case Automation Toolset at Age 14},
  author={Denney, Ewen and Pai, Ganesh},
  booktitle={AIAA SCITECH 2026 Forum},
  pages={1251},
  year={2026}
}

@article{maksimov2019survey,
  title={A survey of tool-supported assurance case assessment techniques},
  author={Maksimov, Mike and Kokaly, Sahar and Chechik, Marsha},
  journal={ACM Computing Surveys (CSUR)},
  volume={52},
  number={5},
  pages={1--34},
  year={2019},
  publisher={ACM New York, NY, USA}
}

@inproceedings{odu2025smartgsn,
  title={SmartGSN: an online tool to semi-automatically manage assurance cases},
  author={Odu, Oluwafemi and Beltr{\'a}n, Daniel M{\'e}ndez and Guti{\'e}rrez, Emiliano Berrones and Belle, Alvine B and Yu, Gerhard and Sherafat, Melika},
  booktitle={International Conference on Computer Safety, Reliability, and Security},
  pages={3--17},
  year={2025},
  organization={Springer}
}

@article{di2024mmint,
  title={MMINT-a: a framework for model-based safety assurance},
  author={Di Sandro, Alessio and Murphy, Logan and Viger, Torin and Chechik, Marsha},
  journal={Science of Computer Programming},
  volume={231},
  pages={103004},
  year={2024},
  publisher={Elsevier}
}

@inproceedings{herd2026subjective,
  title={A Subjective Logic-based method for runtime confidence updates in safety arguments},
  author={Herd, Benjamin and Kelly, Jessica and Heinemann, Clarissa and Zacchi, Jo{\~a}o Vitor},
  booktitle={Proceedings of the 41st ACM/SIGAPP Symposium on Applied Computing},
  pages={1478--1485},
  year={2026}
}

@article{mohamad2023cascade,
  title={Cascade: An asset-driven approach to build security assurance cases for automotive systems},
  author={Mohamad, Mazen and Jolak, Rodi and Askerdal, {\"O}rjan and Stegh{\"o}fer, Jan-Philipp and Scandariato, Riccardo},
  journal={ACM transactions on cyber-physical systems},
  volume={7},
  number={1},
  pages={1--26},
  year={2023},
  publisher={ACM New York, NY}
}

@article{hou2024large,
  title={Large language models for software engineering: A systematic literature review},
  author={Hou, Xinyi and Zhao, Yanjie and Liu, Yue and Yang, Zhou and Wang, Kailong and Li, Li and Luo, Xiapu and Lo, David and Grundy, John and Wang, Haoyu},
  journal={ACM Transactions on Software Engineering and Methodology},
  volume={33},
  number={8},
  pages={1--79},
  year={2024},
  publisher={ACM New York, NY}
}

@article{he2025llm,
  title={Llm-based multi-agent systems for software engineering: Literature review, vision, and the road ahead},
  author={He, Junda and Treude, Christoph and Lo, David},
  journal={ACM Transactions on Software Engineering and Methodology},
  volume={34},
  number={5},
  pages={1--30},
  year={2025},
  publisher={ACM New York, NY}
}

@inproceedings{belle2025preliminary,
  title={A Preliminary Systematic Review on LLM-Based System Assurance: Is Generative AI All You Need?},
  author={Belle, Alvine Boaye and Yu, Gerhard},
  booktitle={2025 IEEE International Conference on Collaborative Advances in Software and COmputiNg (CASCON)},
  pages={589--590},
  year={2025},
  organization={IEEE}
}

@article{viger2026evaluating,
  title={Evaluating AI-supported eliminative argumentation for developing reliable assurance cases},
  author={Viger, Torin and Murphy, Logan and Diemert, Simon and Menghi, Claudio and Babikian, Aren A and Joyce, Jeff and Di Sandro, Alessio and Anwari, Naweed and Cyffka, Erin and Chechik, Marsha},
  journal={Empirical Software Engineering},
  volume={31},
  number={4},
  pages={101},
  year={2026},
  publisher={Springer}
}

@inproceedings{carlan2020enhancing,
  title={Enhancing state-of-the-art safety case patterns to support change impact analysis},
  author={Carlan, Carmen and Gallina, Barbara},
  booktitle={30th European safety and reliability conference},
  volume={136},
  year={2020}
}

@inproceedings{annable2025principled,
  title={Principled Safety Assurance Arguments},
  author={Annable, Nicholas and Lawford, Mark and Paige, Richard F and Wassyng, Alan},
  booktitle={International Conference on Computer Safety, Reliability, and Security},
  pages={18--32},
  year={2025},
  organization={Springer}
}

@inproceedings{shahandashti2024using,
  title={Using GPT-4 turbo to automatically identify defeaters in assurance cases},
  author={Shahandashti, Kimya Khakzad and Belle, Alvine Boaye and Mohajer, Mohammad Mahdi and Odu, Oluwafemi and Lethbridge, Timothy C and Hemmati, Hadi and Wang, Song},
  booktitle={2024 IEEE 32nd International Requirements Engineering Conference Workshops (REW)},
  pages={46--56},
  year={2024},
  organization={IEEE}
}

@inproceedings{viger2023supporting,
  title={Supporting assurance case development using generative ai},
  author={Viger, Torin and Murphy, Logan and Diemert, Simon and Menghi, Claudio and Di, Alessio and Chechik, Marsha},
  booktitle={SAFECOMP 2023, position paper},
  year={2023}
}

@article{hawkins2021guidance,
  title={Guidance on the assurance of machine learning in autonomous systems (AMLAS)},
  author={Hawkins, Richard and Paterson, Colin and Picardi, Chiara and Jia, Yan and Calinescu, Radu and Habli, Ibrahim},
  journal={arXiv preprint arXiv:2102.01564},
  year={2021}
}

@inproceedings{gauerhof2020assuring,
  title={Assuring the safety of machine learning for pedestrian detection at crossings},
  author={Gauerhof, Lydia and Hawkins, Richard and Picardi, Chiara and Paterson, Colin and Hagiwara, Yuki and Habli, Ibrahim},
  booktitle={International Conference on Computer Safety, Reliability, and Security},
  pages={197--212},
  year={2020},
  organization={Springer}
}

@inproceedings{nearchou2023assurance,
  title={An assurance case driven development paradigm for autonomous vehicles: An f1tenth racing car case study},
  author={Nearchou, Ioannis and Rafalko, Lance and Phillips, Ryan and Anderson, Matthew and Shen, Wuwei and Drager, Steve},
  booktitle={2023 IEEE/ACIS 21st International Conference on Software Engineering Research, Management and Applications (SERA)},
  pages={156--161},
  year={2023},
  organization={IEEE}
}

@inproceedings{sorokin2024towards,
  title={Towards continuous assurance case creation for ads with the evidential tool bus},
  author={Sorokin, Lev and Bouchekir, Radouane and Beyene, Tewodros A and Liao, Brian Hsuan-Cheng and Molin, Adam},
  booktitle={European Dependable Computing Conference},
  pages={49--61},
  year={2024},
  organization={Springer}
}

@article{borg2023ergo,
  title={Ergo, SMIRK is safe: a safety case for a machine learning component in a pedestrian automatic emergency brake system},
  author={Borg, Markus and Henriksson, Jens and Socha, Kasper and Lennartsson, Olof and Sonnsj{\"o} L{\"o}negren, Elias and Bui, Thanh and Tomaszewski, Piotr and Sathyamoorthy, Sankar Raman and Brink, Sebastian and Helali Moghadam, Mahshid},
  journal={Software quality journal},
  volume={31},
  number={2},
  pages={335--403},
  year={2023},
  publisher={Springer}
}

@inproceedings{sivakumar2024design,
  title={Design of the safety case of the reinforcement learning-enabled component of a quanser autonomous vehicle},
  author={Sivakumar, Mithila and Belle, Alvine B and Shan, Jinjun and Odu, Oluwafemi and Yuan, Mingfeng},
  booktitle={2024 IEEE 32nd International Requirements Engineering Conference Workshops (REW)},
  pages={57--67},
  year={2024},
  organization={IEEE}
}

@article{odu2025automatic,
  title={Automatic instantiation of assurance cases from patterns using large language models},
  author={Odu, Oluwafemi and Belle, Alvine B and Wang, Song and Kpodjedo, Segla and Lethbridge, Timothy C and Hemmati, Hadi},
  journal={Journal of Systems and Software},
  volume={222},
  pages={112353},
  year={2025},
  publisher={Elsevier}
}

@inproceedings{viger2024ai,
  title={AI-supported eliminative argumentation: Practical experience generating defeaters to increase confidence in assurance cases},
  author={Viger, Torin and Murphy, Logan and Diemert, Simon and Menghi, Claudio and Joyce, Jeff and Di Sandro, Alessio and Chechik, Marsha},
  booktitle={2024 IEEE 35th International Symposium on Software Reliability Engineering (ISSRE)},
  pages={284--294},
  year={2024},
  organization={IEEE}
}

@article{yu2026llms,
  title={LLMs as judges: Toward the LLM-assisted review of GSN-compliant assurance cases},
  author={Yu, Gerhard and Sivakumar, Mithila and Belle, Alvine B and Ghari, Soude and Wang, Song and Lethbridge, Timothy C},
  journal={Journal of Systems and Software},
  pages={113036},
  year={2026},
  publisher={Elsevier}
}

@inproceedings{diemert2025balancing,
  title={Balancing the risks and benefits of using large language models to support assurance case development},
  author={Diemert, Simon and Cyffka, Erin and Anwari, Naweed and Foster, Olivia and Viger, Torin and Millet, Laure and Joyce, Jeffrey},
  booktitle={International Conference on Computer Safety, Reliability, and Security},
  pages={209--225},
  year={2025},
  organization={Springer}
}

@inproceedings{gohar2024codefeater,
  title={Codefeater: Using llms to find defeaters in assurance cases},
  author={Gohar, Usman and Hunter, Michael C and Lutz, Robyn R and Cohen, Myra B},
  booktitle={Proceedings of the 39th IEEE/ACM International Conference on Automated Software Engineering},
  pages={2262--2267},
  year={2024}
}

@article{ballingall2025safety,
  title={Safety assurance for automated systems in transport: A collective case study of real-world fatal crashes},
  author={Ballingall, Stuart and Sarvi, Majid and Sweatman, Peter},
  journal={Journal of Safety Research},
  volume={92},
  pages={27--39},
  year={2025},
  publisher={Elsevier}
}

@article{ahlbrecht2025system,
  title={A system-theoretic assurance framework for safety-driven systems engineering: A. Ahlbrecht et al.},
  author={Ahlbrecht, Alexander and Sprockhoff, Jasper and Durak, Umut},
  journal={Software and Systems Modeling},
  volume={24},
  number={1},
  pages={253--270},
  year={2025},
  publisher={Springer}
}

@inproceedings{odu2025llm,
  title={LLM-Based Safety Case Generation for Baidu Apollo: Are We there Yet?},
  author={Odu, Oluwafemi and Belle, Alvine Boaye and Wang, Song},
  booktitle={2025 IEEE/ACM 4th International Conference on AI Engineering--Software Engineering for AI (CAIN)},
  pages={222--233},
  year={2025},
  organization={IEEE}
}

@inproceedings{fujiwara2025llm,
  title={LLM-Based Automated Mitigation and Assurance Case Generation Against Threats to AI Systems},
  author={Fujiwara, Yuya and Tuchida, Takuma and Miyata, Rikuho and Washizaki, Hironori and Ubayashi, Naoyasu},
  booktitle={2025 IEEE Conference on Artificial Intelligence (CAI)},
  pages={906--911},
  year={2025},
  organization={IEEE}
}

@article{sivakumar2024prompting,
  title={Prompting GPT--4 to support automatic safety case generation},
  author={Sivakumar, Mithila and Belle, Alvine B and Shan, Jinjun and Shahandashti, Kimya Khakzad},
  journal={Expert Systems with Applications},
  volume={255},
  pages={124653},
  year={2024},
  publisher={Elsevier}
}

@inproceedings{delavara2023assurance,
  title={Assurance of software-intensive medical devices: What about mental harm?},
  author={de La Vara, Jose Luis and Gallina, Barbara and Fern{\'a}ndez-Caballero, Antonio and Molina, Jose Pascual and Garc{\'\i}a, Arturo S and Ayora, Clara},
  booktitle={2023 53rd Annual IEEE/IFIP International Conference on Dependable Systems and Networks-Supplemental Volume (DSN-S)},
  pages={168--172},
  year={2023},
  organization={IEEE}
}

@article{muram2023attest,
  title={ATTEST: Automating the review and update of assurance case arguments},
  author={Muram, Faiz Ul and Javed, Muhammad Atif},
  journal={Journal of systems architecture},
  volume={134},
  pages={102781},
  year={2023},
  publisher={Elsevier}
}

@inproceedings{kelly2007reviewing,
  title={Reviewing assurance arguments-a step-by-step approach},
  author={Kelly, Tim},
  booktitle={Workshop on assurance cases for security-the metrics challenge, dependable systems and networks (DSN)},
  pages={25--88},
  year={2007}
}

@article{koopman2023ul,
  title={Ul 4600: What to include in an autonomous vehicle safety case},
  author={Koopman, Philip},
  journal={Computer},
  volume={56},
  number={05},
  pages={101--104},
  year={2023},
  publisher={IEEE Computer Society}
}

@inproceedings{finnegan2014security,
  title={A security argument pattern for medical device assurance cases},
  author={Finnegan, Anita and McCaffery, Fergal},
  booktitle={2014 IEEE International Symposium on Software Reliability Engineering Workshops},
  pages={220--225},
  year={2014},
  organization={IEEE}
}

@inproceedings{carlan2016integrated,
  title={Integrated formal methods for constructing assurance cases},
  author={C{\^a}rlan, Carmen and Beyene, Tewodros A and Ruess, Harald},
  booktitle={2016 IEEE International Symposium on Software Reliability Engineering Workshops (ISSREW)},
  pages={221--228},
  year={2016},
  organization={IEEE}
}

@article{murugesan2024automating,
  title={Automating semantic analysis of system assurance cases using goal-directed asp},
  author={Murugesan, Anitha and Wong, Isaac and Arias, Joaqu{\'\i}n and Stroud, Robert and Varadarajan, Srivatsan and Salazar, Elmer and Gupta, Gopal and Bloomfield, Robin and Rushby, John},
  journal={Theory and Practice of Logic Programming},
  volume={24},
  number={4},
  pages={805--824},
  year={2024},
  publisher={Cambridge University Press}
}

@inproceedings{ghahremani2024avoiding,
  title={Avoiding confirmation bias in a safety management system (sms)},
  author={Ghahremani, Ehsan and Joyce, Jeff and Lechner, Sid},
  booktitle={2024 Integrated Communications, Navigation and Surveillance Conference (ICNS)},
  pages={1--11},
  year={2024},
  organization={IEEE}
}

@misc{gsnstandard,   
    title = {GSN (Version 3)},   
    url = { https://scsc.uk/gsn},   
    author = {Goal Structuring Notation Standard Working Group, GSN_SWG},   
    year = {2021},   
    note = {Accessed on May 29, 2026} 
}

@article{lewis2020retrieval,
  title={Retrieval-augmented generation for knowledge-intensive nlp tasks},
  author={Lewis, Patrick and Perez, Ethan and Piktus, Aleksandra and Petroni, Fabio and Karpukhin, Vladimir and Goyal, Naman and K{\"u}ttler, Heinrich and Lewis, Mike and Yih, Wen-tau and Rockt{\"a}schel, Tim and others},
  journal={Advances in neural information processing systems},
  volume={33},
  pages={9459--9474},
  year={2020}
}

@article{brown2025systematic,
  title={A systematic literature review of retrieval-augmented generation: Techniques, metrics, and challenges},
  author={Brown, Andrew and Roman, Muhammad and Devereux, Barry},
  journal={arXiv preprint arXiv:2508.06401},
  year={2025}
}

@inbook{kodama2024,
author = {Kodama, Hideaki and Matsuno, Yutaka and Takai, Toshinori and Ota, Hiroshi and Okada, Manabu and Tsuchiya, Tomoyuki},
year = {2024},
month = {09},
pages = {150-165},
title = {A Case Study of Continuous Assurance Argument for Level 4 Automated Driving},
isbn = {978-3-031-68605-4},
doi = {10.1007/978-3-031-68606-1_10}
}

@techreport{weinstock2009towards,
  title={Towards an assurance case practice for medical devices},
  author={Weinstock, Charles B and Goodenough, John B},
  year={2009}
}

@article{yu2026safety,
  title={Safety Case Patterns for VLA-based driving systems: Insights from SimLingo},
  author={Yu, Gerhard and Ishikawa, Fuyuki and Odu, Oluwafemi and Belle, Alvine Boaye},
  journal={arXiv preprint arXiv:2603.16013},
  year={2026}
}

@inproceedings{zeng2026experience,
  title={An Experience Report on LLM-Based Agentic Translation from Android to iOS: Pitfalls and Insights},
  author={Zeng, Zhili and Shahandashti, Kimya Khakzad and Belle, Alvine Boaye and Wang, Song and Jiang, Zhen Ming},
  booktitle={Proceedings of the 2026 IEEE/ACM Third International Conference on AI Foundation Models and Software Engineering},
  pages={6--16},
  year={2026}
}

\end{document}